\documentclass[epl-draft]{epl2}

\usepackage[T1]{fontenc}
\usepackage{graphicx}
\usepackage{dcolumn}
\usepackage{bm}
\usepackage{amsmath}
\usepackage{amssymb}
\usepackage[colorlinks=true]{hyperref}

\begin{document}

\title{Scaling and universality in the kinetic smoothening of interfaces: Application to the analysis of the relaxation of rough vicinal steps of an oxide surface}
\shorttitle{Scaling and universality in the kinetic smoothening of interfaces}
\author{T.T.T. Nguyen \and D. Bonamy \and L. Phan Van \and J. Cousty \and L. Barbier}
\institute{CEA, IRAMIS, SPCSI, 91191 Gif-sur-Yvette, France}

\abstract{Relaxation of initially out-of-equilibrium rough interfaces in presence of thermal noise is investigated using Langevin formalism. During thermal equilibration towards the well-known roughening regime, three scaling regimes observed over three successive ranges of length-scales are evidenced:  thermal roughening (late stage) at small scales, transient smoothening at intermediate scales and remnant of the initial conditions at large scales. A generalization of the Family-Vicsek scaling is found for the smoothening regime. A distinctive feature of the transient smoothening regime resides in the existence of a super-universal exponent, i.e. independent of the considered model. This approach allows interpreting a series of AFM images of sapphire surfaces showing the thermal evolution of initially rough step edges.}


\date{\today}
\maketitle

Roughening of growing surfaces and interfaces is a ubiquitous phenomenon in nature, ranging from wetting front \cite{Ertas94_pre,Moulinet04_pre}, flame propagation \cite{Zhang92_pa,Mullys00_prl}, bacterial growth \cite{Vicsek90_pa}, fluid flows in porous media \cite{Sahimi95_book,Alava04_ap} and fracture\cite{Hansen91_prl,Bouchaud97_jpcm,Bonamy09_jpd}, to cite a few. Extensive theoretical and experimental studies have shown that these very different systems follow morphological scaling properties at large scale that can be classified into few universality classes characterized by the values of scaling exponents \cite{Barabasi95_book}. In most cases, the dynamics can be described by a growth Langevin-type equation:
\begin{equation} 
\frac{\partial h}{\partial t} = \Phi ( \vec{\nabla} h) + \xi(\vec{x},t)
\label{langevin}
\end{equation}
\noindent where $h(\vec{x},t)$ is the height of the interface at substrate location $\vec{x}$ and time $t$, $\Phi(\vec{\nabla} h)$ is a function that defines a particular model, and $\xi$ is a noise term. In the steady regime, where the interface fluctuations are statistically time invariant, the morphological scaling features of the surface can be characterized by computing the height-height correlation function $G(\Delta \vec{x})=\left<(h(\vec{x}+\Delta \vec{x})-h(\vec{x}))^2\right>$ that scales as $G(|\Delta x|) \propto |\Delta x|^\zeta$. The value of the so-called roughness exponent $\zeta$ is found to be the same in very different systems and depends on the system dimension, the noise correlation, and the symmetry of $\Phi(\vec{\nabla} h)$, only \cite{Barabasi95_book}.
 
Kinetic roughening from initially flat conditions has been widely investigated, both experimentally \cite{Balankin06_prl,Omi06_prl,Morel08_pre,Cordoba09_prl} and theoretically \cite{Family85_jpa,Krup94_prl,Lopez99_prl,Escudero08_prl}. The time evolution $G(\Delta x,t)$ of the height-height correlation function was shown to obey the dynamic scaling:  
\begin{equation}
\begin{array} {l}
    G^{Rough}(\Delta x,t)= t^ {2\beta_R} f(\Delta x/t^{1/z}) \\
\\
\mathrm{where} \quad	f(u) = \left\{
\begin{array}{l l} 
f_0 \times (u/u_0)^{2\zeta} & $if u$ \ll u_0  \\
f_0 & $if u$ \gg u_0
\end{array}
\right.
\end{array}
\label{Groughapprox}
\end{equation}

\noindent where $\beta_R$ and $z$ refer to the growth exponent and the dynamic exponent, respectively. Two distinct scaling are defined: (i) Family-Vicsek scaling where the three exponents are related through $z=\zeta/\beta_R$ \cite{Family85_jpa} and (ii) anomalous scaling where this last relation is not fulfilled \cite{Lopez99_prl}. While the constants $f_0$ and $u_0$ depend on the precise form of $\Phi$ and $\xi$, the exponents $\zeta$, $z$ and $\beta_R$ (or $\zeta$ and $z$ only for Family-Vicsek scaling) allow to characterize entirely the universality class of the considered model. 

Beyond this simple situation of perfectly smooth initial conditions (hardly relevant in experiments), initiating the growth (or smoothening) from various non equilibrium initial conditions induces transient regimes that, as shown here, may exhibit also universal scaling properties, the analysis of which allows extracting valuable information on the system. In this respect, Bustingorry et al. \cite{Bustingorry06_prl,Bustingorry07a_jsm,Bustingorry07b_jsm} have recently considered the case of an interface driven out from its initial thermal equilibrium by modifying suddenly the working "temperature", i.e. the variance of the noise term in Eq. \ref{langevin}. They evidenced a complex aging dynamics reminiscent of glassy systems. Here, we consider the transient flattening regime from an initially highly rough surfaces, and derive an analytical solution for its dynamics. A new universal dynamic scaling applying far from equilibrium is evidenced. The predictions are then compared to experimental AFM images showing the thermal relaxation of initially rough sapphire surfaces. The time evolution of the step height correlation function can be perfectly reproduced and values of both the step stiffness and the atom hopping rate are deduced.

\begin{figure}
\begin{center}
\includegraphics[width=0.9\columnwidth]{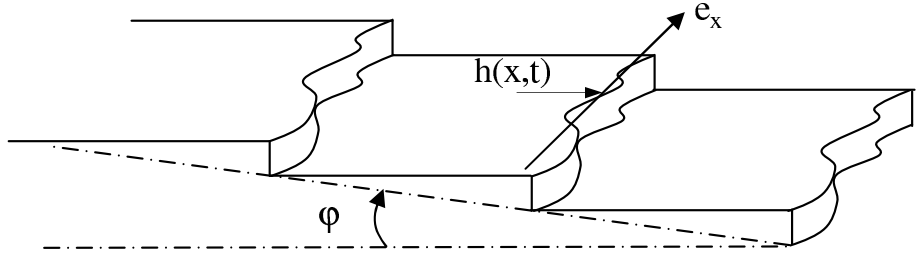}
\caption{Sketch and notations introduced in the text: The frame $\{\vec{e}_x,\vec{e}_y\}$ is chosen so that $\vec{e}_x$ and $\vec{e}_y$ are perpendicular and parallel to the mean step edges respectively, $\varphi$ denotes the angle between terrace and the mean surface plane.}
\label{sketch}
\end{center}
\end{figure}

Specifically, we consider a vicinal surface as depicted in Fig. \ref{sketch} where (i) the step density - controlled by the vicinal angle - is low enough so that step-step interaction can be neglected and (ii) matter is not conserved along the steps. This situation corresponds to the one investigated experimentally and will allow for {\em quantitative} comparisons between model's predictions and experiments. However, as will be discussed, the out-of-equilibrium scalings obtained thereafter appear to be generic and can be extended to other smoothening dynamics. 

For the considered case (no step-step interaction, no matter conservation along steps), the time evolution of step profiles can be described by Langevin Eq. \ref{langevin} with \cite{Bartelt92_ss,Masson94_ss,Barbier96_ss,Legoff99_ss}:
\begin{equation}
\Phi(\nabla h)=\nu\frac{\partial^2 h}{\partial x^2}, \quad \langle \xi(x,t) \xi(x',t') \rangle =D \delta(x-x')\delta(t-t')
\label{EW}
\end{equation}
\noindent where $\nu=\Gamma \eta/kT$ and $D=2\Gamma$, with $\Gamma$ the atom hopping rate, $\eta$ the stiffness of the step edge, $T$ the temperature, and $\delta(u)$ the usual delta function. This equation is classically referred to as the one-dimensional Edward-Wilkinson (1d-EW) equation. To solve this equation analytically, we first discretize the step profile: $h_{n_x}(t)=h(x=n_x a_x,t)$ ($n_x=\{-N/2,...,N/2-1\}$ with $a_x$ the lattice constant) and call $\tilde h_q(t)=(1/\sqrt{N}){\sum_{n_x}} h_{n_y}(t) \exp(-2\pi i q \frac{n_x}{N})$ the $x-$ discrete Fourier transform of $h$. The solution of Eq. \ref{langevin} writes: 

\begin{equation}
\tilde h_q(t)=\exp\left(-\frac{4\pi^2}{a_x^2 N^2}\nu q^2 t\right) \left( \tilde h_q(0)+ \int _0^t \tilde \eta_q(u) \exp\left(\frac{4\pi^2}{a_x^2 N^2}\nu q^2 u\right)du \right)
\end{equation}
 
 
\noindent where $\tilde \xi_q(t)$ and $\tilde h_q(0)$ are the $x$-Fourier transform of $\xi(x,t)$ and of the initial step profile $h(x,0)$, respectively. The complete time evolution of the spatial correlation function $G(\Delta x,t)$ is deduced:

\begin{eqnarray}
\lefteqn {G(\Delta x,t)=} \nonumber\\
 & & \frac{2 D a_x N}{\pi^2 \nu } \sum_{q=1}^{N/2} \frac{1}{q^2} \left( 1 - \exp\left(-\frac{8\pi^2}{a_x^2 N^2}\nu q^2 t\right) \right) \left(1 - \cos( 2\pi \Delta x\frac{q}{a_x N})\right)\nonumber\\
 & & + \frac{4}{N} \sum_{q=1}^{N/2} |\tilde h_q(0)|^2  \exp\left(-\frac{8\pi^2}{a_x^2 N^2}\nu q^2 t\right) \left( 1-\cos( 2\pi \Delta x\frac{q}{a_x N}) \right),\nonumber\\
\label{Eqcorrelation}
\end{eqnarray} 
\noindent that separates into the sum of two terms:

\begin{equation}
G=G^{Rough}(\Delta x,t)+G^{Smooth}(\Delta x,t)
\label{Gtotal} 
\end{equation}

\begin{figure}
\begin{center}
\includegraphics[width=0.9\columnwidth]{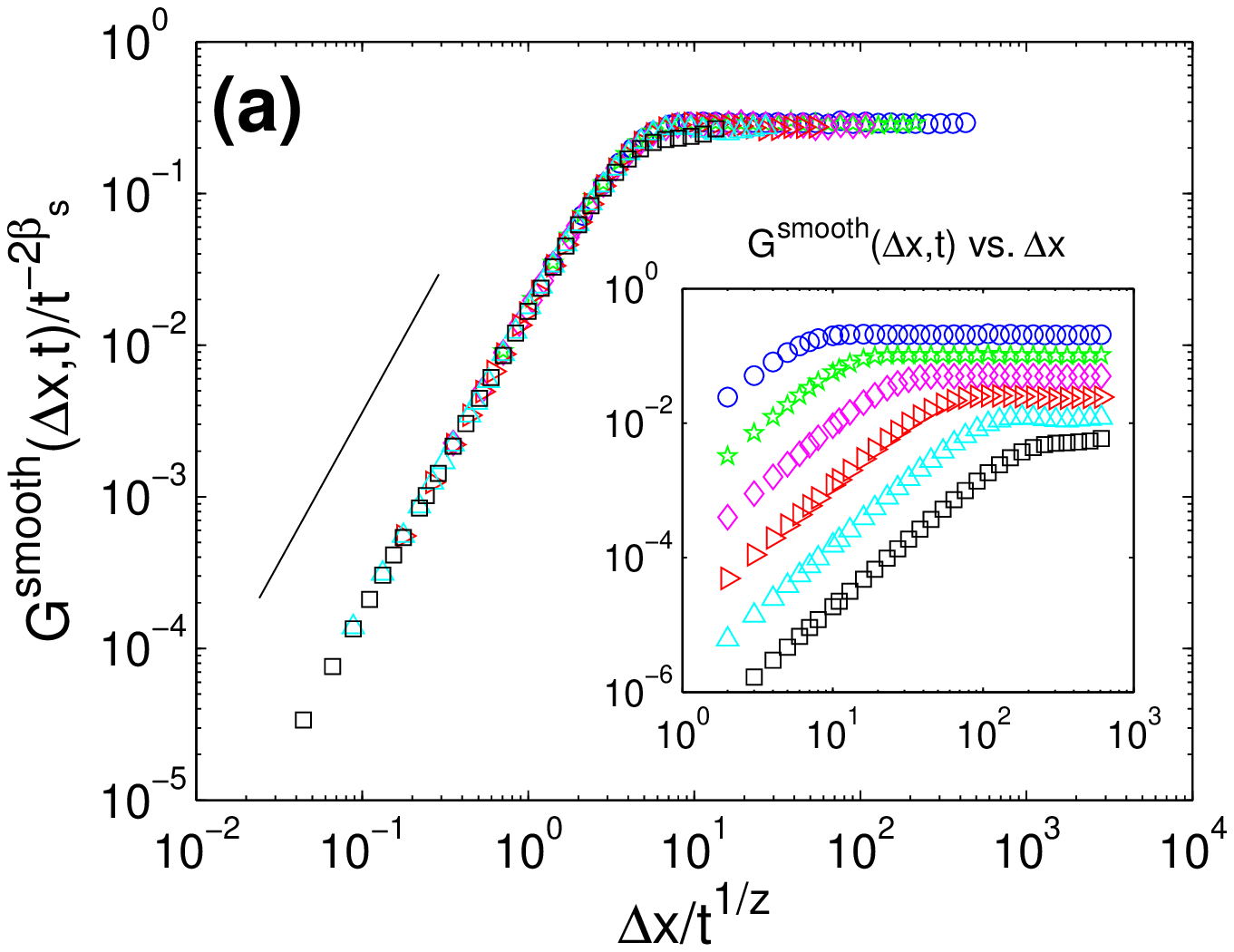}
\includegraphics[width=0.9\columnwidth]{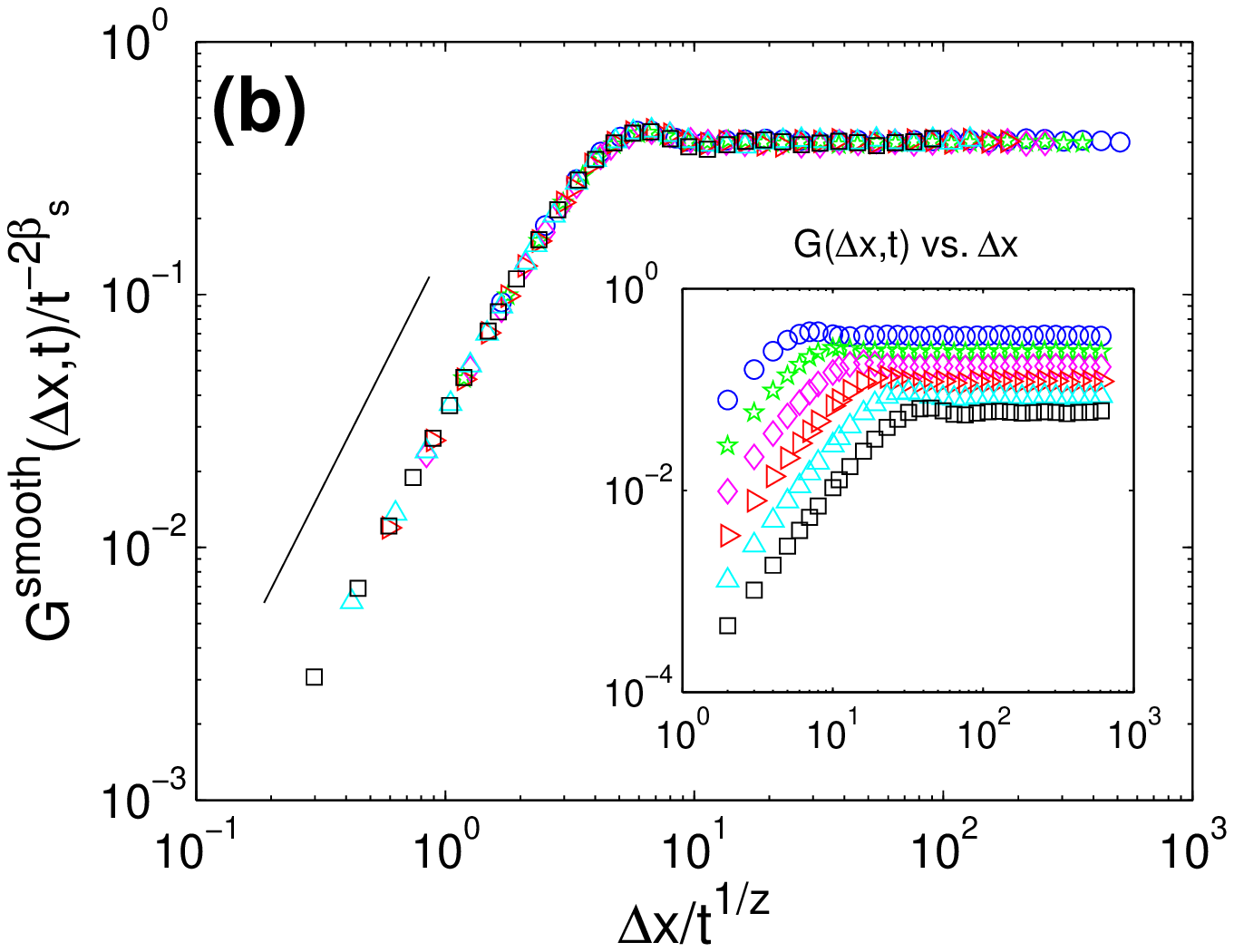}
\caption{Dynamic scaling of the height-height correlation function $G^{Smooth}(\Delta x,t)$ in the out-of-equilibrium smoothening regime for (a) EW equation (Eq. \ref{EW}) with $\nu=1$, and (b) Langevin equation incorporating surface diffusion and conservative noise (Eq. \ref{LCC4}) with $K=1$. The scaling exponents are found to be $\{z=2,\beta_S=1/2z=1/4\}$ in (a), and $\{z=4,\beta_S=1/2z=1/8\}$ in (b). In both cases, the axis are logarithmic and the straight plain lines correspond to power-laws with an exponent of 2. The initial condition $h(t=0)$ is a rough uncorrelated profile of uniform distribution, zero average and $\sigma_0^2=1$ variance. The various symbols correspond to successive time steps, namely $t=2$ ($o$), $t=8$ ($\star$), $t=32$ ($\diamond$), $t=128$ ($\rhd$), $t=512$ ($\triangle$), $t=2048$ ($\Box$).}
\label{simulation}
\end{center}
\end{figure}

\noindent The first term, $G^{Rough}(\Delta x,t)$, describes the roughening of an initially flat step under the action of the thermal noise $\xi$ and takes the dynamic scaling given by Eq. \ref{Groughapprox} with 1d-EW exponents $\{\zeta=1/2,\beta_R=1/4,z=\zeta/\beta_R=2\}$ and parameters $\{f_0\propto D\nu^{-1/2},u_0 \propto \nu^{1/2}\}$. The second term, $G^{Smooth}(\Delta x,t)$, describes the smoothening of the initial profile $h(y,t=0)$ due to the effective line tension $\nu$ in absence of thermal fluctuations. It is then the competition between the disorder in the initial front morphology and the elastic term that sets universal dynamic scaling in this latter case. Let us consider the case of an initially uncorrelated profile with zero average and $\sigma_0^2$ variance, the dynamics of which is described by the 1d-EW Eq. \ref{EW}. Then, $G^{Smooth}(\Delta x,t)$ is found to take the following scaling form :
\begin{equation}
\begin{array} {l}
    G^{Smooth}(\Delta x,t)= t^ {-2\beta_S} g(\Delta x/t^{1/z}) \\
\\
\mathrm{where} \quad	g(v) = \left\{
\begin{array}{l l} 
g_0 \times (v/v_0)^2 & $if v$ \ll v_0  \\
g_0 & $if v$ \gg v_0
\end{array}
\right.,
\end{array}
\label{Gsmoothapprox}
\end{equation}
 \noindent where the scaling exponents are $\beta_S=1/4$, $z=2$, and the parameters $v_0$ and $g_0$ are given by $v_0 \propto \nu^{1/2}$ and $g_0 \propto \sigma_0^2 a_x\nu^{-1/2}$, respectively. Note the existence of an exponent of 2 at small scales, instead of $2\zeta$ in the case kinetic roughening starting from initially flat conditions (Eq. \ref{Groughapprox}). This dynamic scaling was directly confronted to direct simulation: Starting at $t=0$ with an initial uncorrelated random front $h_0$ of $1024$ points with zero average and unit variance, the time evolution of the profile $h(x,t)$ is computed over $t=8192\un{units}$ by solving Eq. \ref{EW} using (i) a finite difference scheme (time step $\delta t = 0.1$) and (ii) periodic boundary conditions ($h(0)=h(L)$).
The time evolution of the height-height correlation function $G(\Delta x,t)$ is then computed and averaged over 50 noise realizations for the initial profile $h_0$. It is found to obey perfectly the dynamic smoothening scaling given by Eq. \ref{Gsmoothapprox} with $\beta_S=1/4$ and $z=2$ (Fig. \ref{simulation}a), as expected.    

Similar analytical development holds for any linear growth model, with conserved or not noise $\xi$, and yields Equation \ref{Gsmoothapprox}.  The two scaling exponents $\beta_S$ and $z$ are found to be related through: 
\begin{equation}
\beta_S=\frac{1}{2z}
\label{ExponentRelation}
\end{equation}
\noindent This relation is analogue to Family-Viscek's one, $\beta_R=\zeta/z$, that intervenes in kinetic roughening from initially flat conditions (see Eq. \ref{Groughapprox}). For instance, let us consider the situation of an initially rough profile that relaxes by diffusion along itself in presence of a conservative noise. Its dynamics is then described by a Langevin growth equation (Eq. \ref{langevin}) with \cite{Wolf90_epl}:  
\begin{equation}
\Phi(\nabla h)=-K\frac{\partial^4 h}{\partial x^4}, \,\, \langle \xi(x,t) \xi(x',t') \rangle =D \frac{\partial}{\partial x}\delta(x-x')\delta(t-t'),
\label{LCC4}
\end{equation}
\noindent As presented in fig. \ref{simulation}b, the numerical solutions of this stochastic equation are found to obey the kinetic smoothening scaling given by Eqs. \ref{Gsmoothapprox} and \ref{ExponentRelation} with the dynamic exponent $z=4$ expected for linear conservative dynamics and conservative noise \cite{Barabasi95_book}.

Scaling and relation given by Eqs. \ref{Gsmoothapprox} and \ref{ExponentRelation} are also conjectured to hold in presence of a non-linear term, like in KPZ equation for instance. In this latter case, the dynamic exponent $z$ that intervenes within the out-of-equilibrium smoothening regime (Eqs. \ref{Gsmoothapprox} and \ref{ExponentRelation}) can be different from that in the standard Family Viseck roughening scaling starting from initially flat conditions (Eq. \ref{Gsmoothapprox}) \cite{note}.

\begin{figure}
\begin{center}
\includegraphics[width=0.9\columnwidth]{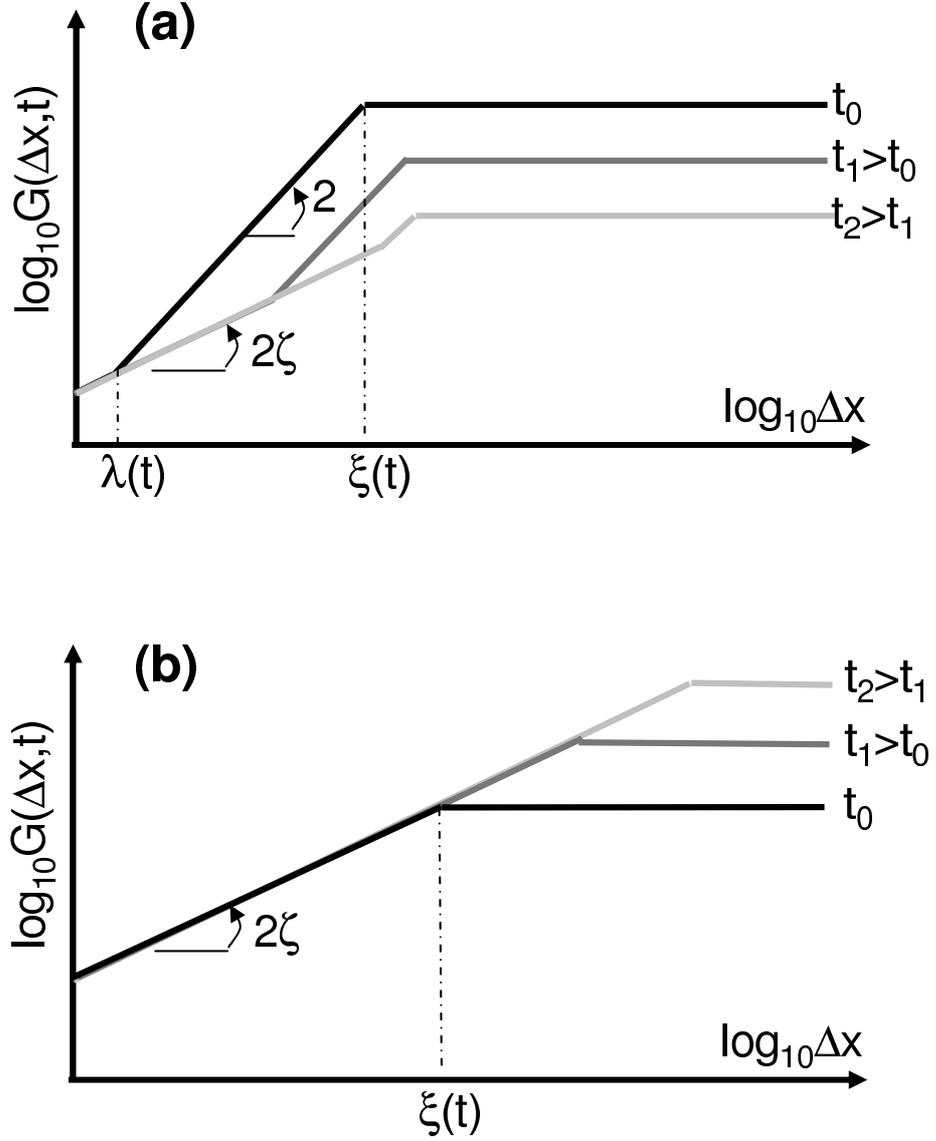}
\caption{Schematic illustration of the spatial correlation function describing the time evolution of a step profile starting from an initially uncorrelated rough profile $h(x,t=0)$ with zero average and $\sigma_0^2$  variance. (a) Smoothing regime observed for small times (see Eq. \ref{equ9}). (b) Roughening regime observed at large times (see Eq. \ref{equ10}).}
\label{Gsketch}
\end{center}
\end{figure}

Back to linear growth models, from the forms of $G^{Rough}(\Delta x,t)$ and $G^{Smooth}(\Delta x,t)$, - given by Eq. \ref{Groughapprox} and Eq. \ref{Gsmoothapprox} respectively -, the global behaviour of $G(\Delta x,t)$ is deduced \cite{note}. It is sketched in Fig. \ref{Gsketch}. Depending on the time $t$, two cases can be distinguished:
\begin{itemize}
\item[-] For small time $t \ll  t_\times$, the variation of the spatial correlation function $G(\Delta x,t)$ can be decomposed into three regimes. At small scales, thermal equilibrium is reached and $G^{Rough}(\Delta x,t)$ is the dominant term in $G(\Delta x,t)$. At medium and large scales, $G(\Delta x,t)$ results from the smoothening of the initial roughness and $G^{Smooth}(\Delta x,t)$ dominates:

\begin{eqnarray}
& \Delta x \ll \lambda(t) & G(\Delta x,t) \propto \Delta x^{2\zeta} \nonumber\\
& \lambda(t) \ll \Delta x \ll \xi(t) &  G(\Delta x,t) \propto t^{-2(\beta_S+1/z)}\Delta x^2 \nonumber\\
& \Delta x \gg \xi(t) & G(\Delta x,t) \propto t^{-2\beta_S},
\label{equ9}
\end{eqnarray}

\noindent with $\lambda(t) \propto t^{(\beta_S+1/z)/(1-\zeta)}$  and $\xi(t) \propto t^{1/z}$.

\item[-] For large time $t \gg t_\times$, the influence of the initial conditions $h(x,t=0)$ is not seen anymore. The spatial correlation function $G(\Delta x,t)$ is then given by $G^{Rough}(\Delta x,t)$ for all $\Delta x$ with two power-law regimes:

\begin{eqnarray}
& \Delta x \ll \xi(t) \quad & G(\Delta x,t) \propto \Delta x^{2\zeta} \nonumber\\
& \Delta x \gg \xi(t) \quad & G(\Delta x,t) \propto t^{2{\beta_R}},
\label{equ10}
\end{eqnarray}

\noindent with $\xi(t) \propto t^{1/z}$.
\end{itemize}
The crossover $t_\times$ between these two regimes is given by:

\begin{equation}
t_\times = \left( \frac{g_0 v_0^2}{f_0 u_0^2} \right)^{1/2(\beta_R+\beta_S)}
\label{equ11}
\end{equation}

Eqs. \ref{equ9}, \ref{equ10} and \ref{equ11} allow to describe entirely the dynamic scaling of a one-dimensional interface described by a Langevin equation (Eq. \ref{langevin}) with any linear growth model $\Phi$ and conserved or not noise $\xi$. In the case of a vicinal step described by the 1d-EW Eq. \ref{EW}, $t_\times$ can be related to $\sigma_0$, $D$ and $\nu$: $t_\times = \sigma_0^2 a_x /D$.

To illustrate the validity and richness of the above approach, it is now applied to analyse AFM observations during the smoothening of vicinal surfaces of oxide surfaces. The samples studied here consist in square slabs of sapphire (Le Rubis S.A., 10mm side, 0.2mm thick) with a surface oriented close to the  $(1,\bar{1},0,2)$ plane (misorientation $\simeq 0.06^\circ$). They present important technological interest as substrate for nanostructures, magnetic thin layers or giant magnetoresistance devices \cite{Fujii89_jap,Eerenstein02_prl,Ramos07_prb}. These surfaces are first chemo-mechanically polished to an optical grade, and carefully cleaned in an ultrasonic bath. As a result, one gets an out-of-equilibrium rough vicinal surface. A series of annealing in air (essential to maintain the surface stoichiometry) at constant temperature for increasing cumulated durations are then performed in an oven. After each annealing step, the sample is cooled down to room temperature (cooling rate of 30K/min) and the vicinal surface is imaged via a Molecular Imaging Pico+ AFM in contact mode with gold coated Si$_3$N$_4$ cantilevers ($0.58\un{N/m}$ stiffness). Special attention was paid so that our setup allows for an accurate and reproducible positioning of the AFM tip onto the surface and allows to image {\it the very same} area after each annealing treatment (see e.g. \cite{Nguyen08_ss} for details). This prevents dispersion due to local variations of the surface roughness.

Depending on the temperature $T$, two regimes can be observed. At low temperature $973\un{K} \leq T \leq 1173 \un{K}$, the evolution of the surface morphology is governed by the coarsening of 2D islands through anisotropic Ostwald ripening during the experimental time. This regime was studied in a previous paper \cite{Nguyen08_ss} and will not be discussed further. At high temperature $1173\un{K} \leq T \leq 1473 \un{K}$, all the islands shrink up and overhangs in steps disappear rapidly, in less than one hour for $T=1173\un{K}$. After this initial regime, the  evolution of the surface morphology is governed by the smoothening of the vicinal steps. 

\begin{figure}
\begin{center}
\includegraphics[width=0.95\columnwidth]{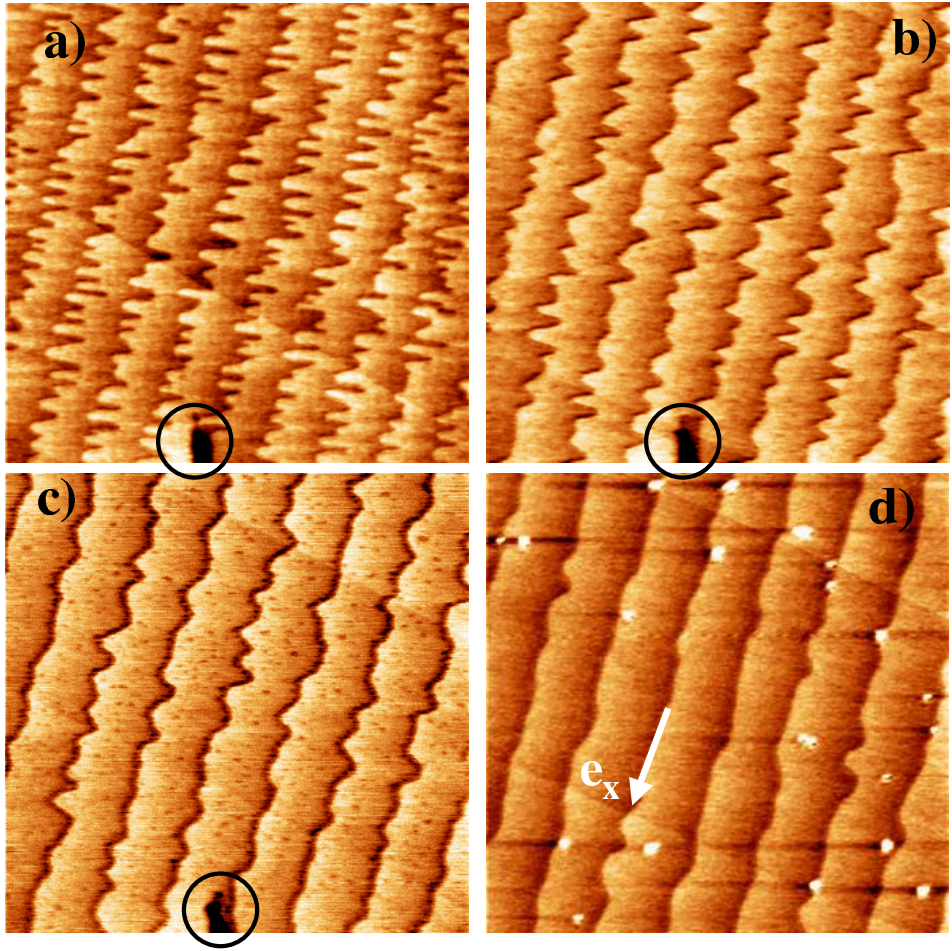}
\caption{$5 \times 5~\mathrm{\mu m}^2$ contact mode AFM images of the topography of a $(1\bar{1}02)$ vicinal surface of sapphire after a) 5400s, b) 12600s, c) 23400s et d) 37800s  of annealing at 1273K. The axis $e_x$ coincides with the $[\bar{1},1,0,1]$ direction. Note the defect (black circles) in (a)-(c) that proves the accurate repositioning of the AFM tip between two successive annealing processes.}
\label{AFM}
\end{center}
\end{figure}

AFM typical images of the vicinal surface taken after cumulated annealing time are presented in Fig. \ref{AFM}. The meandering lines are the steps defining the terraces. Contrast enhancement, image analysis and edge detection allow measuring $h(m,x,t)$ and then computing $G(\Delta x,t)=\langle (h(m,x+\Delta x,t)-h(m,x,t))^2\rangle_{m,x}$ where the average has been performed over all edges of a given image to improve the statistics. Note that the mean distance between two successive steps remains constant and pretty large ($\simeq 750\un{nm}$). This suggests that the morphological fluctuations of a given step are independent from the neighbouring ones, as assumed in the 1-d EW description (Eq. \ref{langevin} and \ref{EW}).

\begin{figure}
\begin{center}
\includegraphics[width=0.9\columnwidth]{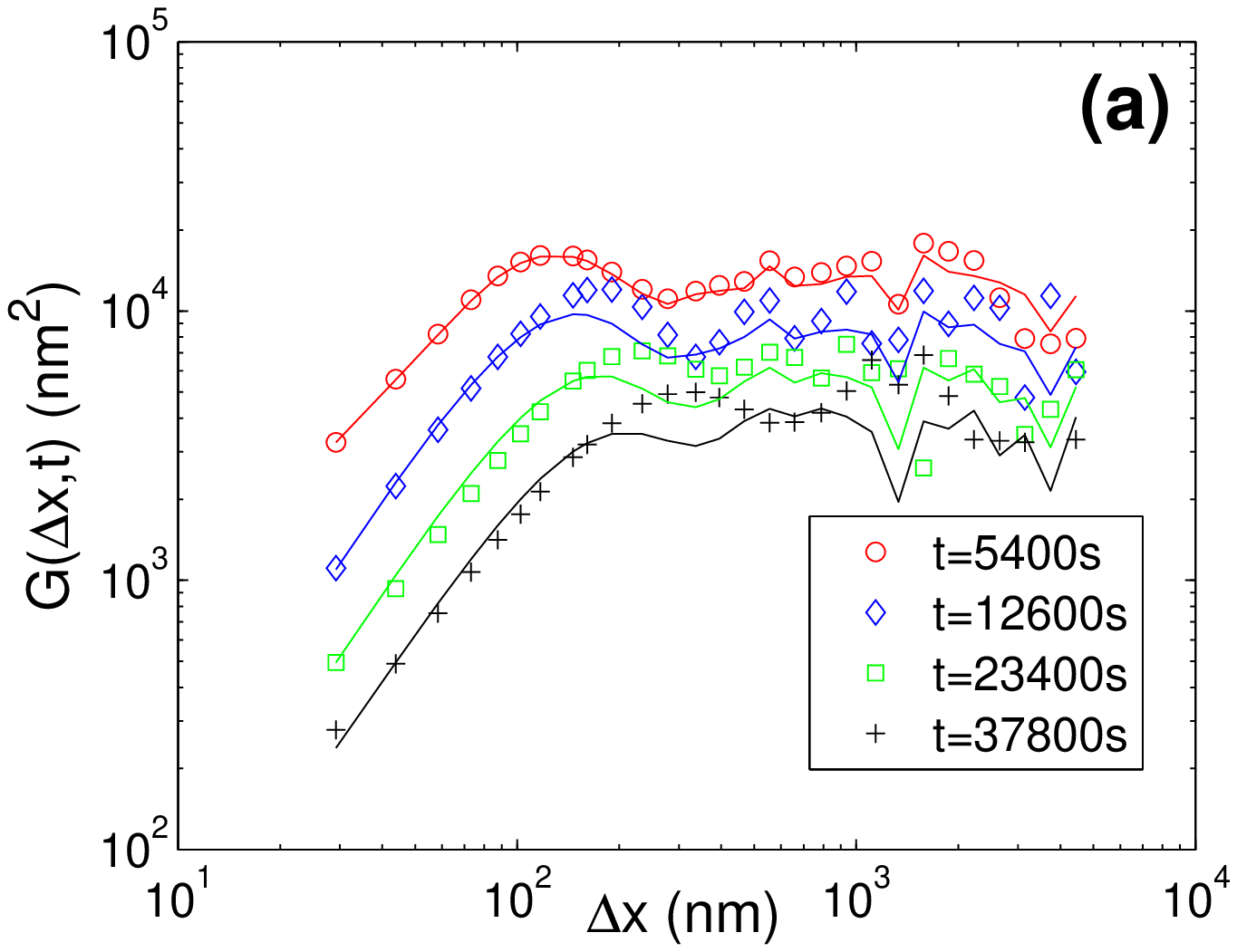}
\includegraphics[width=0.9\columnwidth]{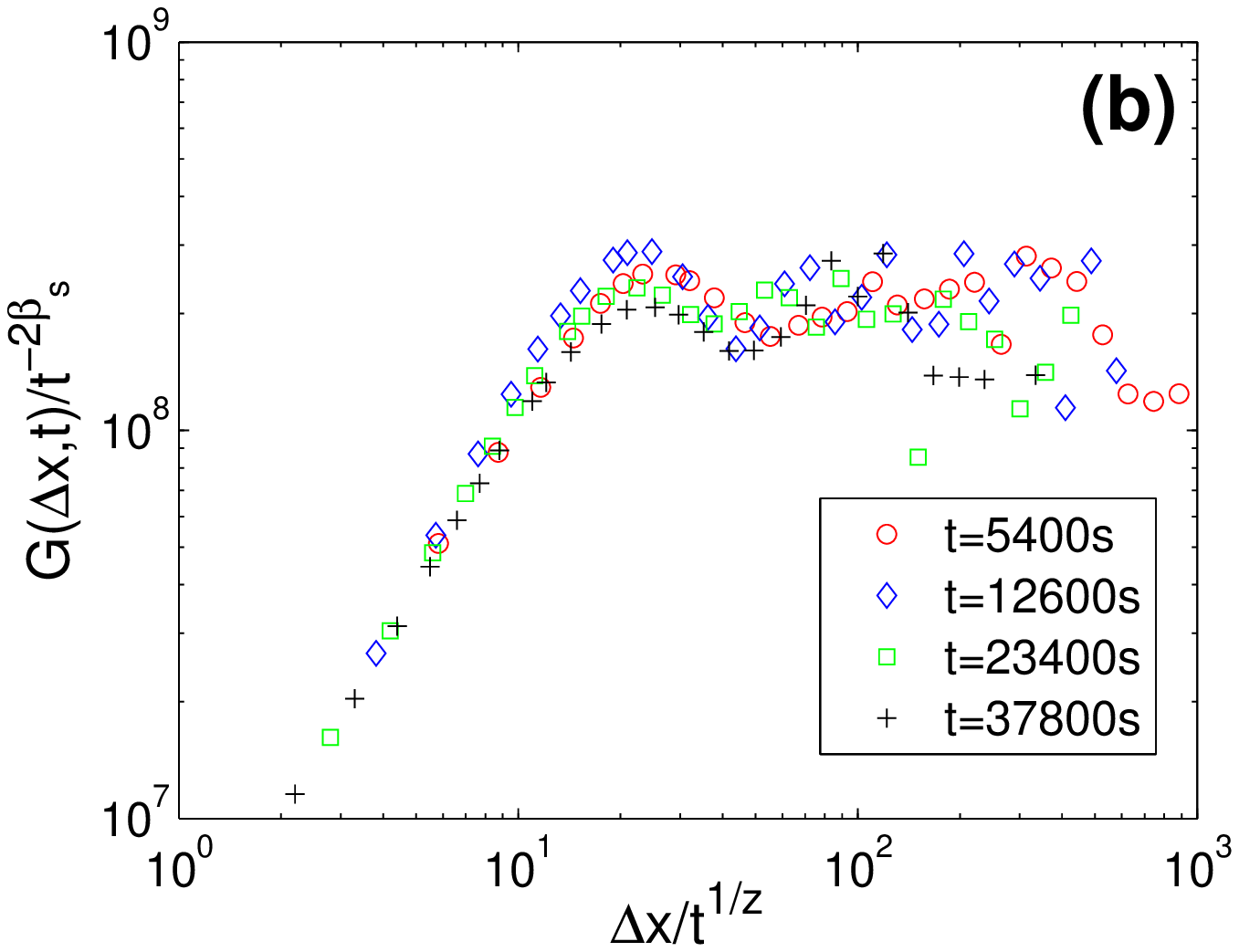}
\caption{(a) Spatial correlation function $G(\Delta x, t)$ of the step edges of the $\mathrm{(1\bar{1}02)}$ alumina surface for various cumulative durations $t$ of annealing at $T=1273\un{K}$. The axis are logarithmic. The symbols correspond to experiments while the continuous lines corresponds to the analytical expression given Eq. \ref{Eqcorrelation} with $\nu=7.2 \times 10^{-2}\un{nm}^{2}.\mathrm{s}^{-1}$ and $D=5.3\times 10^{-2}\un{nm}^{3}.\mathrm{s}^{-1}$. (b) Collapse using Eq. \ref{Gsmoothapprox} with scaling exponents $\{z=2,\beta_S=1/2z=1/2\}$.}
\label{Figcorrelation}
\end{center}
\end{figure}

The experimental correlation functions are plotted in Fig. \ref{Figcorrelation}. They exhibit the scaling given by Eq. \ref{Gsmoothapprox} (Inset of Fig. \ref{Figcorrelation}) expected in the far-from-equilibrium smoothening regime. Initial conditions are set by the image taken at $t_0=5400\un{s}$, the step profiles of which give the initial Fourier amplitudes $|\tilde{h}_q|^2(t_0)$. The subsequent experimental correlation functions are then fitted successfully using the analytical expression (Eq. \ref{Eqcorrelation}) (continuous lines in Fig. \ref{Figcorrelation}). This provides a severe test in favor of the present extension of Langevin formalism to far out-of-equilibrium systems. 

Note that the linear regime expected at thermal equilibrium is not apparent after more than $10$h annealing. From the fitted values $\nu = 7.2 \times 10^{-2}\un{nm}^2.\mathrm{s}^{-1}$ and $D = 5.3 \times 10^{-2}\un{nm}^3.\mathrm{s}^{-1}$ (Fig. \ref{Figcorrelation}), the initial roughness $\sigma_0 \simeq 100 \un{nm}$, and from the pixel size $a_x = 14.6 \un{nm}$ of the images, one estimates the time $t_\times$ to pass from the out-of-equilibrium smoothening regime to the equilibrium roughening regime: $t_\times = \sigma_0^2 a_x/D \simeq 32$ days. This makes this last regime difficult to observe experimentally and emphasizes all the importance of the out-of-equilibrium extension of the Langevin formalism to determine the physical parameters in oxide surfaces. This also means that the mechanisms of diffusion on the sapphire surface could be identified before those responsible for the atom mobility in bulk alumina \cite{Heuer08_jecs}

In 1d Langevin linear growth models, the dynamic exponent $z$ depends usually whether or not the dynamics and the noise are conservative (see \cite{Barabasi95_book} for review). In particular, $z=4$ for linear conservative dynamics and conservative noise that is expected at the onset of active diffusion \cite{Legoff03_ss} and commonly observed in metallic surfaces \cite{Giesen01_pss}. Such a regime is not observed for the $(1,\bar{1},0,2)$ sapphire surfaces investigated here. Two origins of these non-conservative processes can be invoked: (i) The channeled structure of terraces which was shown to influence the Ostwald ripening at low $T$ \cite{Nguyen08_ss}; These channels parallel to $[\bar{1},1,0,1]$ , i.e. roughly perpendicular to the step edge, make the mobility along the step edge difficult. As a result, atoms emitted from the step are diffusing onto the terraces then captured on one step at a different site (detachment/attachment mechanism); (ii) The exchange of oxygen atoms, during annealing, with the surrounding atmosphere, evidenced by a different morphology of terraces after UHV annealing at 1273K \cite{Nguyen08_phd}. This makes the detailed extension of the present work to higher dimensions of high interest. It would also be important to see to which extent the scaling proposed here (Eqs. \ref{Gsmoothapprox}, \ref{equ9} and \ref{equ10}) holds in presence of non-linear terms in the growth models \cite{note}. Work in this direction is currently under progress. 

In conclusion, we have derived here an analytical solution for the dynamics of an initially out-of-equilibrium surface, described by a linear Langevin-like growth equation. For an initially uncorrelated rough profile, the correlation functions during the smoothening phase exhibit a universal dynamic scaling (Eq. \ref{Gsmoothapprox}): 
In the intermediate regime, the correlation function scales with the distance with an exponent of 2 while at large distances, the correlation function remains a plateau. The time scaling exponents for the crossover is $1/z$  and $-2{\beta_S}$ for the plateau level, where $z$ is dynamic exponent and $\beta_S$ relates to $z$ through $\beta_S=1/2z$.
As for other critical systems, these two scaling exponents are function of the correlation range of the disorder. It is then worth noting that in the smoothening process, the relevant disorder is set by the initial profile. In other words, {\em the initial conditions will define the universality class}.  For example, it is interesting to consider the case on an initially self-affine profile $h(x,t=0)$ characterized by an initial roughness exponent $\zeta_0$ . This situation is encountered in the evolution of a profile at thermal equilibrium $T_0$ after a thermal quenching at a new temperature $T$ considered theoretically in \cite{Bustingorry06_prl,Bustingorry07a_jsm,Bustingorry07b_jsm} or starting from the various available morphologies reached after growth \cite{Omi05_prl,Frisch06_prl,Hamouda08_ss}. 
Then the universality class of the smoothening is a function of $\zeta_0$, and, e.g. $z=2$ and $\beta_S=-\zeta_0/2$ in a smoothening process described by the 1d-EW equation (Eq. \ref{EW}).  As a consequence, the study of kinetic smoothening in experimental systems where the initial morphology can be varied opens interesting perspectives in the understanding of critical phenomena. 

\begin{acknowledgments}
The authors thank the invaluable technical support of C. Lubin, F. Thoyer and S. Foucquart. Interesting discussions with S. Bustingorry are also gratefully acknowledged.
\end{acknowledgments}


\begin{thebibliography}{100}
\bibitem{note}
Numerical simulations of KPZ equations reveal that the kinetic smoothening regime can be described using the scaling given by Eq. \ref{Gsmoothapprox} with an effective exponent $z=2$, and not the dynamic exponent $z=3/2$ that intervenes in the Family-Viseck scaling that characterizes the kinetic roughening at equilibrium. To be published.
\end{thebibliography}
\end{document}